\documentclass[twocolumn,floats,showpacs,amsmath,amssymb,pre]{revtex4}
\usepackage{graphicx}
\begin{document}

\title{Compressibility of solid helium}
\author{Carlos P. Herrero}
\affiliation{Instituto de Ciencia de Materiales de Madrid,
         Consejo Superior de Investigaciones Cient\'{\i}ficas (CSIC), 
         Campus de Cantoblanco, 28049 Madrid, Spain }
\date{\today}

\begin{abstract}
The compressibility of solid helium ($^3$He and $^4$He) in the hcp and 
fcc phases has been studied by path-integral Monte Carlo.
Simulations were carried out in both canonical ($NVT$) and  
isothermal-isobaric ($NPT$) ensembles at temperatures between 
10 and 300 K, showing consistent results in both ensembles.
For pressures between 4 and 10 GPa, the bulk modulus $B$ is found
to decrease by about 10\%, when temperature increases from the 
low-temperature limit to the melting temperature.
The isotopic effect on the bulk modulus of helium crystals has 
been quantified in a wide range of parameters.
At 25 K and pressures on the order of 1 GPa, the relative difference
between $^3$He and $^4$He amounts to about 2\%.
The thermal expansion has been also quantified from results obtained
in both $NPT$ and $NVT$ simulations.
\end{abstract}


% 67.80.B- Solid <sup>4</sup>He
% 67.80.de Structure, lattice dynamics and sound
% 62.50.-p High-pressure effects in solids and liquids
% 02.70.Uu Applications of Monte Carlo methods

\maketitle

\section{Introduction}

In the last decades there has been a continuous progress in the
study of different types of substances under extreme conditions of pressure
and temperature, thus enlarging appreciably the experimentally accessible
region of phase diagrams \cite{yo91}.
In particular, the influence of controlled hydrostatic pressure on
structural, thermodynamic, and electronic properties of various kinds
of solids has been deeply studied.
Pressures on the order of tens of GPa can now be routinely applied 
to real materials \cite{lo93,he89,sh01,er06}.

Solid helium, in spite of having been studied for many years,
has a broad interest in condensed matter physics because of its
peculiar character of ``quantum solid''.
In particular, its zero-point vibrational energy and associated anharmonic
effects are markedly larger than in most known solids \cite{gl76}.
In addition, its electronic simplicity allows one to carry out 
detailed studies, that would be enormously difficult for other materials
\cite{ce95,na05,ca08}. 
The interest on the behavior of solids under high pressures has
been also focused on solid helium.
Thus, diamond-anvil-cell and shock-wave experiments have allowed
to study the equation of state (EOS) of solid $^4$He up to
pressures on the order of 50 GPa \cite{po86,ma88,lo93}.
In last years, the effect of pressure on heavier rare-gas solids
has also been of interest for both experimentalists \cite{sh01,er02} and
theorists \cite{ne00,ii01,de02,ts02}.

 The Feynman path-integral formulation of statistical mechanics 
\cite{fe72,kl90} is well suited  to study thermodynamic 
properties of solids at temperatures lower than their Debye temperature
$\Theta_D$, where the quantum character of the atomic nuclei becomes
important.
In particular, the combination of path integrals with computer simulation
methods, such as Monte Carlo or molecular dynamics, has revealed as a powerful
technique to carry out quantitative and nonperturbative studies of 
many-body quantum systems at finite temperatures. This has allowed to study
several properties of solids further than the usual harmonic or quasiharmonic
approximations \cite{ce95}.

 The path-integral Monte Carlo (PIMC) method has been 
used to study several properties of solid helium 
\cite{ce95,ce96,ba89,bo94,ch01,he06,he07}, 
as well as heavier rare-gas solids \cite{cu93,mu95,ch02,ne02,he02,he05}.
For helium, in particular, this method has predicted kinetic-energy 
values \cite{ce96} and Debye-Waller factors \cite{dr00} in good agreement with 
data derived from experiments \cite{ar03,ve03}.
PIMC simulations have been also employed to study the isotopic shift
in the helium melting curve \cite{ba89,bo94}.
The EOS of solid helium at $T$ = 0 has been studied by diffusion Monte Carlo in 
a wide density range \cite{mo00,ca08}, as well as at finite temperatures
by using PIMC simulations with several interatomic potentials \cite{ch01}. 

In last years, there has been a debate on the existence of supersolidity
in $^4$He at temperatures lower than 1 K \cite{ki04,ri07,cl07,ce04}. 
This debate is still open, but is
out of the scope of this paper, since we consider here solid helium at
temperatures higher than 10 K, where quantum exchange effects between atomic
nuclei are not relevant.

In this paper, we study the compressibility of solid $^3$He and $^4$He 
by PIMC simulations. We employ the isothermal-isobaric ($NPT$) ensemble,
which allows us to consider properties of these solids along well-defined 
isobars. For comparison, we present also results of PIMC simulations in
the canonical ($NVT$) ensemble.
By comparing results for $^3$He and $^4$He, we analyse isotopic
effects on the compressibility as a function of pressure.

The paper is organized as follows.  In Sec.\,II, the
computational method is described. In Sec.\,III we present and discuss 
the results, that are given in  several subsections dealing with thermodynamic
consistency, pressure and temperature dependence of the bulk modulus, 
isotopic effects, and thermal expansion.
 Finally, in Sec.\,IV we present the conclusions.

\section{Method}
                                                                                    
Equilibrium properties of solid $^3$He and $^4$He in the face-centred cubic 
(fcc) and hexagonal close-packed (hcp) phases have been
calculated by PIMC simulations. 
The PIMC method relies on an isomorphism between the quantum system under 
consideration and a fictitious classical one, obtained by replacing each
quantum particle by a cyclic chain of $Q$ classical particles
($Q$: Trotter number), connected
by harmonic springs with a temperature-dependent force constant.
This isomorphism appears as a consequence of discretizing the density
matrix along cyclic paths, which is usual in the path-integral formulation
of statistical mechanics \cite{fe72,kl90}.  Details on this computational
method are given elsewhere \cite{ch81,gi88,ce95,no96}.

Helium atoms were considered as quantum particles interacting
through an effective interatomic potential, composed of a two-body
and a three-body part.  For the two-body interaction, we employed
the potential developed by Aziz {\em et al.} \cite{az95}
(the so-called HFD-B3-FCI1 potential). For the three-body
part we took a Bruch-McGee-type potential \cite{br73,lo87},
with the parameters given by Loubeyre \cite{lo87}, but with
parameter $A$ in the attractive exchange interaction rescaled by a
factor 2/3, as suggested in Ref. \cite{bo94}.
This interatomic potential was found earlier to describe well the vibrational
energy and equation-of-state of solid helium in a broad range of pressures
and temperatures \cite{he06}.

Our simulations were based on the so-called ``primitive'' form
of PIMC \cite{ch81,si88}.
We considered explicitly two- and three-body terms in the simulations.
The actual consideration of three-body terms 
did not allow us to use effective forms for the density matrix,
developed to simplify the calculation when only two-body
terms are explicitly considered \cite{bo94}.
Quantum exchange effects between atomic nuclei were not taken into account,
because they are negligible for solid helium at the temperatures and
pressures studied here. (This is expected to be valid assuming the absence of
vacancies and for temperatures higher than the exchange frequency 
$\sim 10^{-6}$ K \cite{ce95}.)
To calculate the energy we have used the ``crude'' estimator, as defined in
Refs. \cite{ch81,si88}.

We have employed both the canonical ($NVT$)
 ensemble and the isothermal-isobaric ($NPT$) ensemble.
Our simulations were performed on supercells of the fcc and hcp
unit cells, including 500 and 432 helium atoms respectively.
To check the convergence of our results with system size, we carried out 
some simulations for other supercell sizes, and found that finite-size 
effects for $N > 400$ atoms are negligible for the quantities
studied here. In particular, changes of the bulk modulus with the size of
the simulation cell were found to be smaller than the statistical error bar
of the values derived from our simulations.

Sampling of the configuration space was carried out by the Metropolis
method at temperatures between 12 K and the
melting temperature at each considered pressure.
 For given temperature and pressure, a typical run consisted
of $10^4$ Monte Carlo steps for system equilibration, followed by $4 \times 10^5$ 
steps for the calculation of ensemble average properties.
Each step included attempts to move every replica of every atom in the
simulation cell. In the $NPT$ simulations, it also included an attempt to 
change the volume. 
To keep roughly constant the accuracy of the computed quantities
at different temperatures, we took a Trotter number $Q$ proportional to
the inverse temperature, so that $Q T$ = 3000~K. 
This means that for solid helium at $T$ = 20 K we had $Q$ = 150,
and a PIMC simulation for $N = 500$ atoms effectively includes 75000 
``classical'' particles. 
More technical details are given in Refs.~\cite{no97,he02,he06}.

In the $NVT$ ensemble the pressure and isothermal bulk modulus 
$B = - V ( {\partial P} / {\partial V} )_T$ can be obtained
from thermal averages of various quantities obtained in PIMC
simulations, as shown in Appendix A.
The derivation is straightforward from the partition function $Z_{NVT}$
for $N$ quantum particles in the canonical ensemble, but is rather tedious
in the case of the bulk modulus,
due to the number of terms appearing in the volume derivatives.  
In particular, $B$ can be obtained from Eqs.~(\ref{bulkm}) and (\ref{dzv2})
in Appendix A.   In the $NPT$ ensemble, 
the isothermal bulk modulus is related to the mean-square fluctuations
of the volume $V$ of the simulation cell by the expression: 
\begin{equation}
      \sigma_V^2 = \frac{V}{B} k_B T   \; ,
\label{dv2}
\end{equation}
as shown in Appendix B.

\section{Results and discussion}

\subsection{Consistency checks}

We have calculated the bulk modulus in the $NPT$ and $NVT$ ensembles.
In general, we prefer the isothermal-isobaric ensemble, so that we
can study solids along well-defined isotherms. $NVT$ simulations are,
however, frequently used in PIMC simulations \cite{ce95,ce96,dr00}, and a
comparison of results obtained in both ensembles seems necessary 
as a consistency check of the method, and in particular for the case
of solid helium. 

We discuss first results obtained in the $NPT$ ensemble.
At constant pressure, the relative fluctuations in the volume,
$\sigma_V / V$, can be found from Eq.~(\ref{dv2}):
\begin{equation}
      \frac{\sigma_V}{V} = \left( \frac{k_B T}{B V} \right)^{\frac12}  \; .
\label{dvv}
\end{equation}
For hcp $^4$He at 25 K we found in PIMC simulations 
$\sigma_V / V = 2.9 \times 10^{-3}$ for $P$ = 2 GPa, 
and a lower value of $1.4 \times 10^{-3}$ for 20 GPa 
(for a simulation cell including $N$ = 432 atoms). 
This means that the product $B V$ increases
as the pressure is raised [see Eq.~(\ref{dvv}], in spite of the
reduction of $V$, indicating that the growth of 
bulk modulus with the pressure dominates in the product $B V$ (see
below).

 For a cubic crystal, the fluctuations in the lattice parameter $a$ 
can be derived from Eq.~(\ref{dv2}) to give:
\begin{equation}
      \sigma_a^2 = \frac{k_B T}{9 L^3 a B}   \; ,
\label{da2}
\end{equation}
where $L^3$ is the number of cubic unit cells in a simulation cell with
side length $L a$.
From Eq.\,(\ref{da2}) one can see that the relative fluctuation
in the lattice parameter, $\sigma_a / a$, scales as $L^{-3/2}$.  
This normalized fluctuation for fcc $^4$He is displayed in Fig. 1 
as a function of $L^{-3/2}$.
Shown are results of PIMC simulations in the $NPT$ ensemble 
at $T$ = 100 K and two pressures: 2 and 7 GPa. 
The linear dependence shown in this figure agrees with
the dependence of $\sigma_a$ on the simulation-cell size given
by Eq.~(\ref{da2}). The different slopes of these lines are
basically due to the change of compressibility with pressure.   

\begin{figure}
\vspace{-2.0cm}
\hspace{-0.5cm}
\includegraphics[width= 9cm]{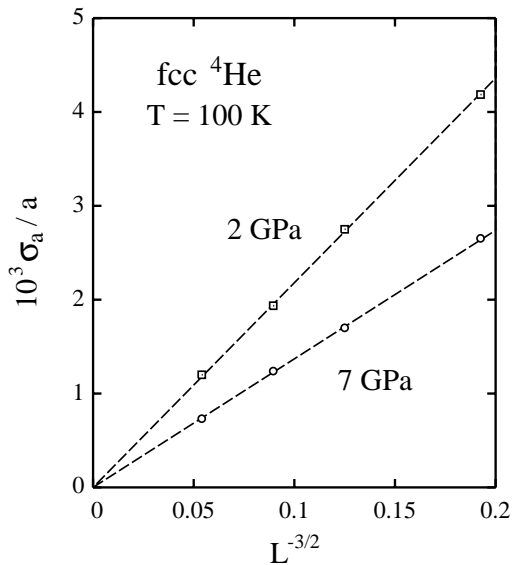}
\vspace{-2.5cm}
\caption{\label{f1}
Relative fluctuation of the lattice parameter, $\sigma_a / a$,
of fcc $^4$He  vs $L^{-3/2}$, for simulation cells of lateral size
$L a$. Shown are results of PIMC simulations for two pressures
(2 and 7 GPa) at $T$ = 100 K, and for $L$ = 3, 4, 5, and 7.
Error bars are less than  the symbol size.
}
\end{figure}

To check the thermodynamic consistency of our simulations in both
$NPT$ and $NVT$ ensembles we have carried out several tests.
The first obvious test consists in taking the crystal volume derived
from isothermal-isobaric simulations, and using it as an input in
$NVT$ simulations at the same temperature. The latter should give the
same pressure [using Eq.~(\ref{pres2})] as that used earlier in the
$NPT$ simulations. As an example, for fcc $^4$He at 25 K and 1 GPa,
we find first a lattice parameter $a$ = 3.6955 \AA, that introduced as an
input in $NVT$ simulations yields a pressure of 1.0002(5) GPa, in good
agreement with the input in the previous $NPT$ simulations.
Going on with the same set of parameters, we can also check the consistency
of the bulk modulus $B$ derived from both types of simulations.
At constant pressure ($P$ = 1 GPa), we find $B_{NPT}$ = 4.50(2) GPa from the
volume fluctuations [see Eq.~(\ref{kappa}) in Appendix B] vs 
$B_{NVT}$ = 4.47(2) GPa derived
at constant volume by using expressions (\ref{bulkm}) and (\ref{dzv2}).

\subsection{Pressure and temperature dependence}

We now consider the pressure dependence of the bulk modulus, as
derived from our PIMC simulations.
This is shown in Fig.~2 for $^4$He at three temperatures.
In this figure, $T$ decreases from top to bottom:
$T$ = 25, 150, and 300 K; open and filled symbols 
correspond to hcp and fcc helium, respectively.
At each considered temperature, we present data for the pressure region
where the solid was stable (or metastable) along the PIMC simulations,
a region that becomes broader as the temperature is lowered.
In addition to the typical rise of $B$ for increasing pressure, we find 
that for a given pressure, $B$ decreases as the temperature 
is raised.
In the pressure range shown in Fig.~2 one also observes a departure
of linearity in the dependence of $B$ on pressure.
By fitting these results to the expression
$B = B_0 + B_0' P + \frac12 B_0'' P^2$, we can find the pressure derivatives
at $P = 0$ ($B_0'$ and $B_0''$). Thus, for $T$ = 25 K we obtain
$B_0$ = 0.47 GPa, $B_0'$ = 4.01, and $B_0''$ = --0.068 GPa$^{-1}$.

\begin{figure}
\vspace{-2.0cm}
\hspace{-0.5cm}
\includegraphics[width= 9cm]{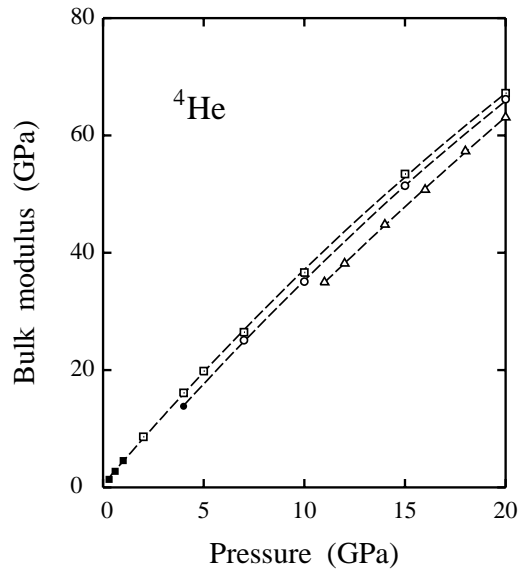}
\vspace{-2.5cm}
\caption{\label{f2}
Isothermal bulk modulus of solid $^4$He as a function of pressure at
three temperatures: 25 K (squares), 150 K (circles), and
300 K (triangles). Open and filled symbols correspond to
hcp and fcc helium, respectively.
Error bars are smaller than the symbol size.
Dashed lines are fits to the data points using the expression:
$B = B_0 + B_0' P + \frac12 B_0'' P^2$.
}
\end{figure}

For the other temperatures shown in Fig.~2 ($T$ = 150 and 300 K), 
an extrapolation of the results to $P=0$ yields
negative values of $B_0$, indicating that
the solid is mechanically unstable at these temperatures for low
pressures. At 25 K and zero pressure, even though the liquid is known to
be the stable phase, the solid can still be metastable, as it has not
yet reached the limit of mechanical stability, or the spinodal line .
Apart from this, the curves $B(P)$ at the considered temperatures are 
rather parallel one to the other in the common stability region.

It is interesting to compare the bulk modulus obtained here for solid helium
with those of heavier rare-gas solids at the same conditions.
For example, at $T$ = 20 K and $P$ = 1 GPa, we find for solid $^4$He:
$B$ = 4.53 GPa, vs 7.20 and 9.43 for Ne and Ar, respectively \cite{he05}.
This is in line with the known result that the bulk modulus increases
with atomic mass for given values of temperature and pressure
\cite{he05}.

Rare-gas solids have been studied in metastable conditions, even at
negative pressures \cite{he03b}. They have been found to be metastable
in PIMC simulations close to the spinodal point, defined as
the point at which the compressibility $\kappa = 1/ B$ diverges (or the 
bulk modulus vanishes). 
Thus, for solid Ne and Ar at 5 K, the spinodal point was
found at $P = -91$ and $-245$ MPa. For solid helium, we could not
approach the spinodal point at any temperature, due to the large quantum
fluctuations that make the solid to be unstable along the simulations.
This happened even at zero pressure and relatively low temperature,
in spite of the fact that the expected bulk modulus at these conditions
is still far from vanishing. For example, at 25 K we find for solid helium
$B_0$ = 0.47 GPa, as derived from extrapolation of the results obtained
at $P >$ 0.2 GPa (see above and Fig.~2), but the solid was unstable in our
simulations at $P <$ 0.2 GPa.

\begin{figure}
\vspace{-2.0cm}
\hspace{-0.5cm}
\includegraphics[width= 9cm]{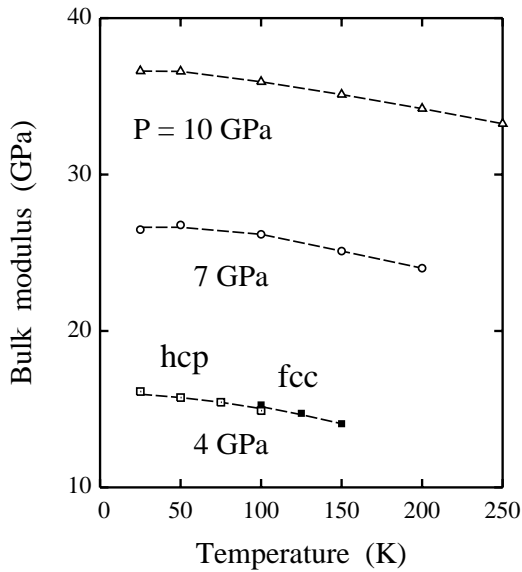}
\vspace{-2.5cm}
\caption{\label{f3}
Isothermal bulk modulus of solid $^4$He as a function of temperature at
three hydrostatic pressures: 4 GPa (squares), 7 GPa (circles), and
10 GPa (triangles).  Open and filled symbols correspond to
hcp and fcc helium, respectively.
Error bars of the simulation results are on the order of the symbol size.
Dashed lines are guides to the eye.
}
\end{figure}

The temperature dependence of the bulk modulus is displayed in Fig.~3
for three different pressures. Here again open and filled symbols
indicate hcp and fcc helium, respectively.
For each pressure under consideration, the plotted data correspond
to the temperature region where we found the solids to be (meta)stable
along the PIMC simulations.
For each pressure, the bulk modulus decreases as the temperature is 
raised, and this decrease is similar for the different pressures.
In fact, the obtained curves $B(T)$ are parallel within the statistical 
error of our simulations.  
In the whole accessible temperature range, the bulk modulus decreases
by 2.1, 2.6, and 3.1 GPa, for $P$ = 4, 7, and 10 GPa, respectively.
These values amount to 13.0, 9.8, and 8.5 \% of the corresponding bulk
modulus at 25 K. 

\begin{figure}
\vspace{-2.0cm}
\hspace{-0.5cm}
\includegraphics[width= 9cm]{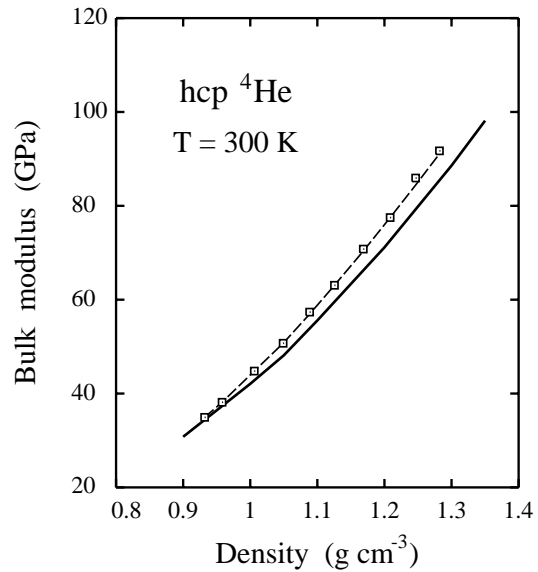}
\vspace{-2.5cm}
\caption{\label{f4}
Isothermal bulk modulus of solid $^4$He versus density.
Open squares represent results of path-integral Monte Carlo simulations
in the $NPT$ ensemble. Error bars are less than the symbol size.
A solid line indicates the bulk modulus derived by
Zha {\em et al.} \cite{zh04} from Brillouin scattering experiments.
The dashed line is a guide to the eye.
}
\end{figure}

To connect with data derived from experiment, we note that
Zha {\em et al.} \cite{zh04} have obtained the bulk modulus of
solid $^4$He from Brillouin scattering measurements. From these results,
they derived the dependence of the isothermal bulk modulus upon material
density at 300 K. 
In Fig.~4 we present results of our PIMC simulations for $^4$He at
300 K (open squares) along with those derived from Brillouin scattering
experiments at room temperature (solid line). At this temperature, 
the density region between 0.95 and 1.25 g cm$^{-3}$ corresponds to a 
pressure between 12 and 29 GPa.
At $\rho < 1 \text{ g cm}^{-3}$, our results coincide within error bars with
those derived from Brillouin scattering, and at higher densities the
bulk modulus derived from PIMC simulations is somewhat larger. 
For example, at density $\rho$ = 1 and 1.25 g/cm$^3$, Zha {\em et al.}
\cite{zh04} obtained $B$ = 42.1 and 79.9 GPa respectively, to
be compared with $B$ = 44 and 86 GPa yielded by our PIMC simulations 
of $^4$He at the same densities and $T$ = 300 K.
This means that at $\rho = 1.25$ g cm$^{-3}$, the difference amounts to 
about 7\%.

\subsection{Isotopic effect}

An interesting point in this context is the influence of the
atomic mass on the solid compressibility. This isotopic effect can
be readily obtained from PIMC simulations, since the mass is an
input parameter in these calculations. Similar isotopic effects have
been studied earlier for the melting curve \cite{ba89,bo94} and molar 
volume \cite{he06,he07} of solid helium.
For a given material, lighter isotopes form more compressible solids,
as a consequence of an increase in the molar volume. 
This is in fact due to a combination of the anharmonicity in the interatomic
potential and zero-point vibrations, which are most important at low 
temperature. Since these quantum vibrations are especially large in the
case of helium, due to its low mass, the isotopic effect on the bulk
modulus is expected to be appreciable.

Thus, one expects the bulk modulus of solid $^4$He to be larger than that of
$^3$He. This is in fact the case, as derived from our PIMC simulations. 
In Fig.~5 we display the difference $\Delta B = B_4 - B_3$ as
a function of pressure at $T$ = 25 K.
For example, at 1 GPa and 25 K we find $B = 4.40$ and $4.50\pm0.01$ GPa for
$^3$He and $^4$He, respectively, and the difference between both 
isotopes amounts to 2.2\%. 
At the same temperature and 7 GPa, we obtained $B$ = 26.20 and
26.59 GPa, which translates to a relative difference of 1.5\%. 
This relative change in bulk modulus associated to the isotopic mass
is appreciable, and slightly larger than that found for the isotopic
effect on the molar volume of helium crystals. Thus, at a pressure of 1 and 
7 GPa and $T$ = 25 K, we find a volume difference of 1.8 and 0.9\%,
when comparing $^3$He and $^4$He \cite{he07}. In general, given a temperature, 
the relative changes in volume and compressibility decrease for
increasing pressure, since the solid behaves as if it was ``more
classical'' \cite{et74,he05}, and isotopic effects (of quantum origin) are 
consequently reduced.

\begin{figure}
\vspace{-2.0cm}
\hspace{-0.5cm}
\includegraphics[width= 9cm]{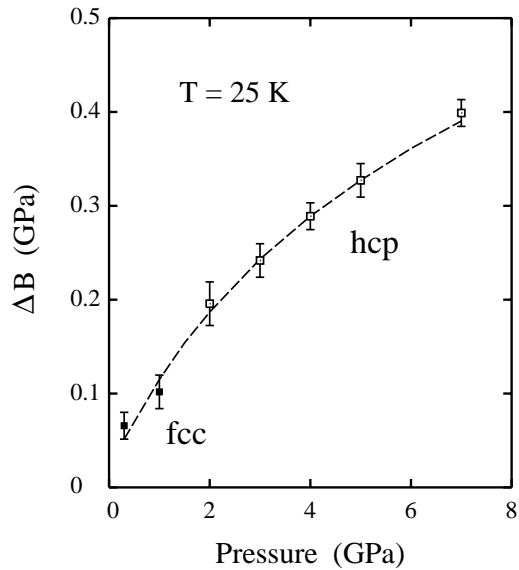}
\vspace{-2.5cm}
\caption{\label{f5}
Isotopic effect on the isothermal bulk modulus as a function of
pressure. Shown is the difference $\Delta B = B_4 - B_3$ between
the bulk modulus of $^4$He and $^3$He at 25 K.
Filled and open symbols represent data for fcc and hcp helium,
respectively.
}
\end{figure}

\subsection{Thermal expansion}

At this point, it is worthwhile comparing
the temperature dependence of the bulk modulus
obtained at fixed volume and fixed pressure, from PIMC simulations in
the $NVT$ and $NPT$ ensembles, respectively.  In Fig. 6 we have plotted 
our results for $B$, as derived in both cases.  On one side,
the simulations in the isothermal-isobaric ensemble were carried out
at a pressure $P$ = 1 GPa, and the results obtained are  represented in Fig. 6 
as circles.    On the other side,
simulations in the canonical ensemble (at various temperatures) 
were performed for the volume obtained in $NPT$ simulations at $T$ = 25 K
and $P$ = 1 GPa.
Results of these $NVT$ simulations are displayed as open squares. 
As expected, $B$ obtained in both types of simulations
agree with each other (within error bars) at $T \le$ 25 K.
At higher temperatures, the $NPT$ simulations yield values of the bulk  
modulus smaller than those derived from $NVT$ simulations.
This is due to the thermal expansion, that is taken into
account in the simulations at constant pressure, but is 
neglected in the constant-volume simulations.
In general, an increase in crystal volume causes a decrease in
the bulk modulus.

\begin{figure}
\vspace{-2.0cm}
\hspace{-0.5cm}
\includegraphics[width= 9cm]{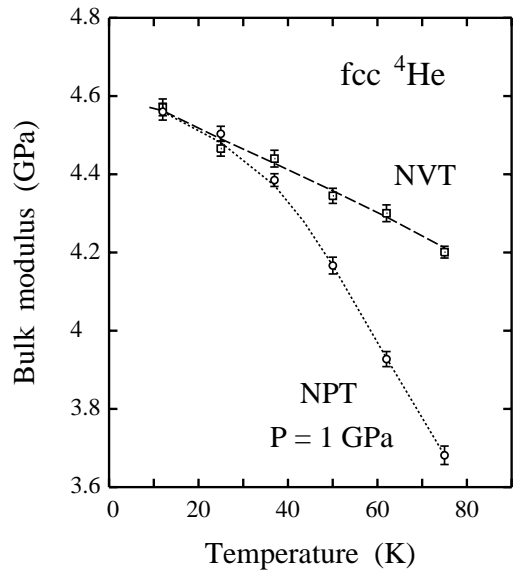}
\vspace{-2.5cm}
\caption{\label{f6}
Bulk modulus vs temperature, as derived from PIMC simulations
at constant volume (squares) and constant pressure (circles).
The simulations in the $NVT$ ensemble were carried out with the
volume obtained at $P$ = 1 GPa and $T$ = 25 K.
Lines are guides to the eye.
}
\end{figure}

In this context, another application of our PIMC simulations consists in 
the calculation of the thermal expansion coefficient 
\begin{equation}
      \alpha =  \frac{1}{V} \left( \frac{\partial V}{\partial T}
       \right)_P   \,   ,
\label{alpha1}
\end{equation}
which can be obtained in the canonical ensemble by using the expression
\cite{ca60}:
\begin{equation}
      \alpha =  \frac{1}{B} \left( \frac{\partial P}{\partial T}
       \right)_V   \;  .
\label{alpha2}
\end{equation}
This equation in fact relates the thermal expansion with the compressibility
through the temperature derivative of the pressure at constant volume.
In connection with Eq.~(\ref{alpha2}), we note that the pressure obtained
in $NVT$ simulations (with a fixed lattice parameter $a$ = 3.6955 \AA) increases
from 1 GPa to 1.3 GPa when temperature rises from 25 to 75 K.
At 25 K, we have $B$ = 4.46 GPa and 
$({\partial P}/{\partial T})_V = 8.87 \times 10^{-4}$ GPa K$^{-1}$, giving 
$\alpha_{NVT} = 1.99 \times 10^{-4}$ K$^{-1}$.  This result is in agreement 
with that found directly from the volume change obtained
in the $NPT$ simulations, which yields 
$\alpha_{NPT} = 2.03 \times 10^{-4} \text{K}^{-1}$. 

\begin{figure}
\vspace{-2.0cm}
\hspace{-0.5cm}
\includegraphics[width= 9cm]{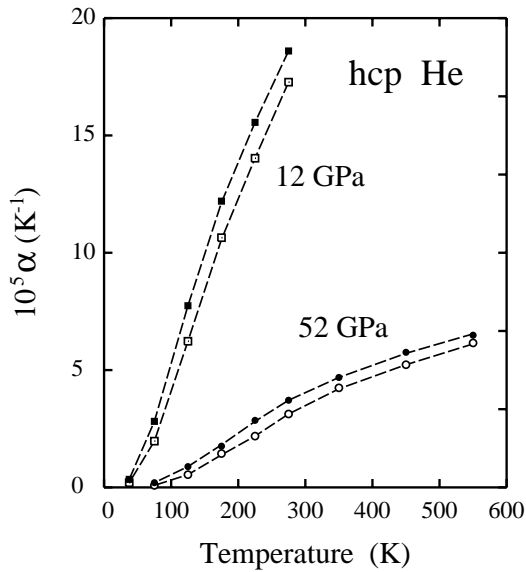}
\vspace{-2.5cm}
\caption{\label{f7}
Temperature dependence of the thermal expansion coefficient $\alpha$
of hcp helium, as derived from PIMC simulations at $P =$ 12 GPa and 52 GPa.
Open and filled symbols correspond to $^3$He and $^4$He, respectively.
Error bars are on the order of the symbol size.
Dashed lines are guides to the eye.
}
\end{figure}

Finally, in Fig. 7  we present results of the thermal expansion coefficient 
$\alpha$ yielded by our simulations in the isothermal-isobaric ensemble at two
(high) pressures. We give values for both $^3$He and $^4$He in the hcp phase.
First, we observe an important decrease in $\alpha$ as the pressure is
raised. For $^4$He at 300 K, we find a reduction in $\alpha$ by a factor
of five, when increasing $P$ from 12 to 52 GPa.
Second, we note that the thermal expansion is smaller for $^3$He than for
$^4$He. This is easy to understand, since the molar volume of solid $^3$He is
larger than that of $^4$He \cite{he06}, and both volumes have to converge
one to the other in the high-temperature (classical) limit.

\vspace{0.2cm}

\section{Conclusions}

Path-integral Monte Carlo simulations in both $NVT$ and $NPT$ ensembles
have been shown to be well suited to analyse the temperature and pressure
dependence of the bulk modulus of solid helium.
In the isothermal-isobaric ensemble, $B$ can be derived from the
volume fluctuations along a simulation run. In the canonical ensemble,
a calculation of $B$ is more elaborate, but  can be carried out from thermal
averages of various intermediate quantities obtained in PIMC simulations. 
Both ensembles $NVT$ and $NPT$ give consistent results for the compressibility 
and thermal expansion of $^3$He and $^4$He in the whole region of
temperatures and pressures considered here.

At a given pressure, the bulk modulus decreases as temperature rises.
For pressures between 4 and 10 GPa, the change in $B$ has been found 
to be on the order of 10\%, when temperature increases from the low-$T$
limit to the melting temperature of the material.

Solid $^3$He is more compressible than $^4$He. At a given $T$, 
the difference between bulk modulus of both solids increases as 
pressure rises, but the relative difference between them decreases. 
This isotopic effect on the compressibility of solid helium
is appreciable in the range of temperatures and pressures studied here.
In fact, at 25 K and pressures on the order of 1 GPa, it amounts to 
about 2\%.

Apart from a precise calculation of the bulk modulus, PIMC simulations 
in the canonical ensemble can be also used to study accurately the thermal 
expansion of the solids under consideration at a given pressure.

\begin{acknowledgments}
The author benefited from useful discussions with R. Ram\'{\i}rez.
This work was supported by Ministerio de Educaci\'on y
Ciencia (Spain) under Contract No. FIS2006-12117-C04-03.  \\
\end{acknowledgments}

\appendix
\section{Pressure and bulk modulus in the $NVT$ ensemble}
\subsection{Pressure}

In the path-integral formalism, the canonical partition function for $N$ 
identical quantum particles obeying Boltzmann statistics (no particle
exchange) can be written as \cite{ch81,gi88}:
\begin{widetext}
\begin{equation}
Z_{NVT} = K  \int \prod_{i=1}^N  \prod_{j=1}^Q   d{\bf r}_{i j}   
     \exp \left( -\beta \sum_{j=1}^Q \left[ C \sum_{i=1}^N 
     ({\bf r}_{i,j+1} - {\bf r}_{ij})^2
   + \frac{1}{Q} \, \Phi({\bf r}_{1j},...,{\bf r}_{Nj}) \right] \right) ,
\end{equation}
\end{widetext}
where ${\bf r}_{ij}$ ($i= 1, ..., N; j = 1, ..., Q$) are Cartesian coordinates
of the $N$ particles at imaginary time $j$, $\beta = (k_B T)^{-1}$, and
$\Phi({\bf r}_{1j},...,{\bf r}_{Nj})$ is the potential energy.
We have defined
\begin{equation}
  K = \frac{1}{N!}  \left( \frac{m Q}{2 \pi \beta \hbar^2} \right)^{3 N Q/2}
\end{equation}
and
\begin{equation}
  C = \frac{m Q}{2 \beta^2 \hbar^2} \; .
\end{equation}
The coordinates are subject to the cyclic condition
${\bf r}_{i,Q+1} = {\bf r}_{i1}$ for all particles $i = 1, ..., N$.
Changing ${\bf r}_{ij}$ to reduced coordinates 
${\bf s}_{ij} = V^{-1/3} {\bf r}_{ij}$, we find:
\begin{equation}
Z_{NVT} = K V^{N Q} \int \prod_{i=1}^N  \prod_{j=1}^Q   d{\bf s}_{i j}
    \exp \left[ -\beta H(\beta,V,\{{\bf s}_{ij}\}) \right]  \; ,
\end{equation}
where we have defined
\begin{eqnarray}
H(\beta,V,\{{\bf s}_{ij}\}) = 
   C V^{2/3} \sum_{i=1}^N \sum_{j=1}^Q ({\bf s}_{i,j+1} - {\bf s}_{ij})^2 + 
       \nonumber \\
       + \frac{1}{Q} \, \sum_{j=1}^Q 
      \Phi(V^{1/3} {\bf s}_{1j},...,V^{1/3} {\bf s}_{Nj})
\end{eqnarray}
and the dependence of $H$ on $\beta$, $V$, and 
the coordinates ${\bf s}_{ij}$ has been explicitly indicated.

The pressure is given by
\begin{equation}
P = \frac{1}{\beta} \frac{1}{Z_{NVT}} \frac{\partial Z_{NVT}}{\partial V} ,
\label{pres1}
\end{equation}
and from the volume derivative of $Z_{NVT}$ one finds:
\begin{equation}
  P =    \frac{N Q}{\beta V} - \frac{2}{3 V} \langle E_{h} \rangle
       - \left< \Phi_V \right>
\label{pres2}
\end{equation}
where $E_h$ refers to the ``harmonic'' energy:
\begin{equation}
  E_h =  C\sum_{i=1}^N  \sum_{j=1}^Q ({\bf r}_{i,j+1} - {\bf r}_{ij})^2
      = C V^{2/3}  \sum_{i=1}^N  \sum_{j=1}^Q
             ({\bf s}_{i,j+1} - {\bf s}_{ij})^2
\end{equation}
and
\begin{equation}
  \Phi_V = \frac{1}{Q} \sum_{j=1}^Q 
         \frac{\partial \Phi(V^{1/3} {\bf s}_{1j},...,V^{1/3} {\bf s}_{Nj})}
         {\partial V} \; .
\end{equation}

\subsection{Bulk modulus}

The isothermal bulk modulus is:
\begin{equation}
B = - V \left( \frac{\partial P}{\partial V} \right)_T
\end{equation}
and substituting Eq.~(\ref{pres1}) for the pressure, we have:
\begin{equation}
B = V \beta P^2 - \frac{V}{\beta Z_{NVT}} \frac{\partial^2 Z_{NVT}}{\partial V^2}
   \; .
\label{bulkm}
\end{equation}
Taking a second volume derivative of $Z_{NVT}$, we find:
\begin{widetext}
\begin{eqnarray}
\label{dzv2}
\frac{1}{\beta Z_{NVT}} \frac{\partial^2 Z_{NVT}}{\partial V^2} & =  &
        \frac{NQ(NQ-1)}{\beta V^2}  
     + \frac{1}{3 V^2} \left( \frac{2}{3} - 4 NQ \right) \left< E_{h} \right> 
     -  \frac{2NQ}{V}  \left< \Phi_V \right>  
     +    \frac{4}{9} \frac{\beta}{V^2}  \left< E_{h}^2 \right>  + \\ 
         \nonumber  \\
 &   +  & \frac{4}{3} \frac{\beta}{V} \left< E_{h} \Phi_V \right> 
     -  \frac{1}{Q} \left< \frac{\partial \Phi_V} {\partial V} \right> 
     +  \beta \left< \Phi_V^2 \right>   \nonumber
\end{eqnarray}
\end{widetext}
Finally, Eqs.~(\ref{bulkm}) and (\ref{dzv2}) give us the bulk modulus $B$ 
at volume $V$ and temperature $T$. 

\section{Compressibility in the $NPT$ ensemble}

We now have the partition function
\begin{equation}
Z_{NPT}  = \int_0^{\infty} dV {\rm e}^{- \beta P V} Z_{NVT} \; .
\end{equation}
To obtain the compressibility, we calculate the pressure derivatives of
$\ln Z_{NPT}$. We find:
\begin{equation}
\frac{\partial \ln Z_{NPT}}{\partial P} =
      \frac{1}{Z_{NPT}} \frac{\partial Z_{NPT}}{\partial P} = 
     - \beta \left< V \right>
\label{npt1}
\end{equation}
and
\begin{eqnarray}
\label{npt2}
\frac{\partial^2 \ln Z_{NPT}}{\partial P^2} & = &
   - \frac{1}{Z_{NPT}^2} \left( \frac{\partial Z_{NPT}}{\partial P} \right)^2 +
     \frac{1}{Z_{NPT}} \frac{\partial^2 Z_{NPT}}{\partial P^2} =  \nonumber  \\
       \nonumber \\
   & = & - \beta^2 \left< V \right>^2 + \beta^2 \langle V^2 \rangle =
    - \beta^2 \sigma_V^2 \; ,  
\end{eqnarray}
where $\sigma_V^2$ is the mean square fluctuation of the volume. 
Moreover, from Eq.~(\ref{npt1}):
\begin{equation}
\frac{\partial^2 \ln Z_{NPT}}{\partial P^2} =
   - \beta  \frac{\partial \langle V \rangle}{\partial P}  \; ,
\label{npt3}
\end{equation}
and from Eqs.~(\ref{npt2}) and (\ref{npt3}), we obtain for the isothermal 
compressibility $\kappa$:
\begin{equation}
 \kappa  =  - \frac{1}{\langle V \rangle}
         \left( \frac{\partial \langle V \rangle}{\partial P} \right)_T
      =  \frac{\beta \sigma_V^2} {\langle V \rangle}  \; .
\label{kappa}
\end{equation}
This is the thermodynamic relation obtained in general in the
isothermal-isobaric ensemble \cite{la80}. Thus, it can be directly used 
to obtain the compressibility from the volume fluctuations in path-integral
simulations, without any further calculations.

\end{document}